\documentclass[aps,prmat,preprint,groupedaddress,floatfix,citeautoscript]{revtex4}
\usepackage{amsmath}
\usepackage{graphicx}
\usepackage{dcolumn}
\usepackage{bm}
\usepackage[mathlines]{lineno}

\newcolumntype{M}[1]{>{\centering\arraybackslash}m{#1}}
\newcolumntype{N}{@{}m{0pt}@{}}

\begin{document}

\title{Fundamental Band Gap and Alignment of Two-Dimensional Semiconductors Explored by Machine Learning
}

\author{Zhen Zhu}
\email%
{zhuzhen@engineering.ucsb.edu}%
\affiliation{Materials Department,
             University of California, Santa Barbara, California 93106, USA}
\author{Baojuan Dong}
\author{Teng Yang}
\email%
{yangteng@imr.ac.cn}%
\author{Zhidong Zhang}
\affiliation{Shenyang National Laboratory
             for Materials Science,
             Institute of Metal Research,
             Chinese Academy of Sciences,
             University of Chinese Academy of Sciences,
             Shenyang 110016, China}

\date{\today} 

\begin{abstract}
Two-dimensional (2D) semiconductors isoelectronic to phosphorene has been drawing much attention recently due to their promising applications for next-generation (opt)electronics. This family of 2D materials contains more than 400 members, including (a) elemental group-V materials, (b) binary III-VII and IV-VI compounds, (c)
ternary III-VI-VII and IV-V-VII compounds, making materials design with targeted functionality unprecedentedly rich and extremely challenging. To shed light on rational functionality design with this family of materials, we systemically explore their fundamental band gaps and alignments using hybrid density functional theory (DFT) in combination with machine learning. First, GGA-PBE and HSE calculations are performed as a reference. We find this family of materials share similar crystalline structures, but possess largely distributed band-gap values ranging approximately from 0 to 8 eV. Then, we apply machine learning methods, including Linear Regression (LR), Random Forest Regression (RFR), and Support Vector Machine Regression (SVR), to build models for prediction of electronic properties. Among these models, SVR is found to have the best performance, yielding the root mean square error (RMSE) less than 0.15~eV for predicted band gaps, VBMs, and CBMs when both PBE results and elemental information are used as features. Thus, we demonstrate machine learning models are universally suitable for screening 2D isoelectronic systems with targeted functionality, and especially valuable for the design of alloys and heterogeneous systems.
\end{abstract}



\maketitle


\section{ Introduction}

Last decade has witnessed the rocketing development of two-dimensional (2D) materials, which find promising applications in next-generation 
electronics and optoelectronics~\cite{NovoselovSci04,PKimNat05,Mak2016,Fiori2014}.
The performance of a 2D electronic device depends sensitively on fundamental electronic properties of the candidate material: a non-zero band gap, proper band edge positions, and high mobility are in general the requisites. In contrary to semi-metallic graphene~\cite{NovoselovSci04,PKimNat05} and low-mobility transition metal dichalcogenides (TMDs)~\cite{Radisavljevic2011,DT221} that fail to deliver good device performance, phosphorene is semiconducting while still maintaining a high hole mobility~\cite{{Li2014},{DT229},{Koenig14},{CastroNeto14},Ling2015}, thereby emerging as a potential candidate for 2D electronics. However, poor chemical stability has limited phosphorene 
for practical applications~\cite{Liu2015}.
To overcome such obstacles, searching for 2D materials with similar electronic properties but better chemical stability is essential.


Recently, high-throughput materials screening has emerged as an effective method to search for materials with targeted functionality~\cite{Curtarolo2013,De2015,Setyawan2010,Hautier2011}. The workflow for materials discovery is separated to different layers: starting with crude and low-precision computations to narrow the candidacy pool, and followed by precise but expensive calculations to identify the candidate materials. The initial materials pool is usually a subset of the ICSD database~\cite{ICSD} with large amount of candidates, resulting in tedious prescreening and large computational efforts. A prescreening method that is both accurate and computationally efficient is greatly desired, where machine learning can play an important role. In combination with density functional theory (DFT), machine learning has
demonstrated valuable applications in functional materials design~\cite{Ward2017}, properties predictions~\cite{Ward2016,Deml2016,Pilania2017,Lee2016}, and many other fields~\cite{Ramprasad2017,Balachandran2015} for traditional bulk materials. It is intriguing to apply such machine learning methods to two-dimensional systems to accelerate materials discovery, which is largely unexplored but fundamentally and technologically important.

Here, we have explored the fundamental band gaps
and band alignments of a group of 2D semiconductors that are isoelectronic to phosphorene
using machine learning techniques in combination with density functional theory.
The methodology is discussed in Sec.~\ref{sec:method}, including details of density
functional calculations (Sec.~\ref{sec:dft_method}) and a brief introduction of machine
learning models (Sec.~\ref{sec:ml_method}). We describe the isoelectronic materials design
method in Sec.~\ref{sec:mat_design}. Following this method, 
more than 400 materials 
are constructed and calculated, including
(a) elemental group-V materials, (b) binary III-VII and IV-VI compounds,
(c) ternary III-VI-VII and IV-V-VII compounds.
Among this family of materials, many have been successfully synthesized~\cite{Ji2016,Zhang2016BlueP} and found special applications in different research fields~\cite{Haleoot2017,Fei2016}.
The richness in electronic properties of these materials is categorized and analyzed
in Sec.~\ref{sec:elec_prop}.
Next, in Sec.~\ref{sec:ml_model}, we apply machine learning methods, including Linear Regression (LR),
Random Forest Regression (RFR), and Support Vector Machine Regression
(SVR), to predict electronic properties for this family of 2D materials. Then we summarize our key findings
in Sec.~\ref{sec:summary}.

\section{Methodology}
\label{sec:method}

\subsection{Computational Details for Density Functional Methods}
\label{sec:dft_method}
All our calculations are based on DFT using projector-augmented waves~\cite{Blochl1994} (PAW)
as implemented in the VASP~\cite{VASP} code. We have used periodic boundary
conditions throughout the study, with monolayer structures represented by a periodic array of
slabs separated by a vacuum region at least 15~{\AA} thick. We used the Perdew-Burke-Ernzerhof
(PBE)~\cite{PBE} exchange-correlation functional for initial structure optimization based on the conjugate gradient
method~\cite{CGmethod} with a 400 eV energy cutoff. All geometries are treated as optimized when none of the residual
Hellmann-Feynman forces exceeded $10^{-2}$~eV/{\AA}. On top of PBE-optimized structures, a single-shot
screened hybrid functional calculation (HSE)~\cite{{Heyd2003},{Heyd2006}} is performed to obtain fundamental
band gaps and band alignments of the material. We have used standard values for the
mixing parameter (0.25) and the range-separation parameter (0.2~\AA$^{-1}$). The
reciprocal space was sampled by a grid~\cite{Monkhorst-Pack76} finer than $10{\times}10{\times}1$~$k$-points
in the Brillouin zone of the primitive unit cell.

\subsection{Machine Learning Methods}
\label{sec:ml_method}

The obtained DFT results are then analyzed with machine learning models as implemented in scikit-learn~\cite{sklearn} package.
Relation between target electronic properties and predictors can be established via supervised learning methods.
A good predictive model depends sensitively on choice of regression models, selection of predictors, as well as the quality
of our dataset. For a given data set, it is important to select proper predictors and suitable regression models to achieve
good predictive ability with high accuracy. To achieve this goal in current study, we have selected three different predictor sets, which are different combinations of computed PBE results and fundamental signatures of constituent elements. Then, we utilize a variety of regression methods, including Linear Regressions, Random Forest Regression, and Support Vector Machine Regression, to predict target electronic properties.

In the LR method, the regression coefficients of predictors, $w$, are determined by optimizing
the following cost function $L({\bf w})$: $L({\bf w})=||{\bf y}-{\bf Xw}||^2$. In addition, other LR methods with regularizations, LASSO and Ridge are also used in this study. On top of the ordinary least square linear regression method, LASSO include an additional $\bf L1$ penalty term
$\sum_i ||\alpha w_i||$ in the cost function, while Ridge regression method add a $\bf L2$ regularization term $\sum_i||\alpha w_i||^2$.
These penalty terms can effectively mitigate the overfitting problem especially when the predictor sets are large.

When the relation between the target property and the predictors is not linear, regression methods like RFR and SVR with a non-linear kernel are supposed to capture the nonlinear feature-target relationship. Random Forest is one type of ensemble methods. It grows a number of decision trees via bootstrapping the sample space. For each decision tree, a randomly selected subset of the feature space is used, which can effectively minimize the correlation between different trees. Then, the target value is predicted by majority vote of these trees for classification or averaging the predicted result of each tree in regression problems. Importantly, the Random Forest model is easy to interpret and It can output the relative importance of different features, thereby providing insights
on the elemental signatures that determines targeted electronic properties of materials in present study.

We also use a SVR model with a Radial Basis Function (RBF) kernel to predict calculated electronic properties with
fundamental materials features. The Support Vector Machine model utilizes the kernel trick to map low-dimensional non-separable data to a higher dimension where they can be separated via a hyper-plane. The optimized hyper-plane can be identified by so-called supported vectors. The kernel trick makes it possible to compute the inner product of
the projected data in the higher dimension without specifying the mapping function, which is usually time-consuming or even impossible to specify. SVR uses a hinge-loss function $\sum_i \rm max (0, 1-y_if(x_i))$, which is minimized during the model training process. The RBF kernel used in present work has the form of $\rm K(x_i,x_j)=exp(-\gamma||x_i-x_j||^2)$.

\begin{table*}[t]
\caption{Three sets of predictors used for machine learning models to predict electronic band gaps and band alignments.}
\begin{tabular}{|c|c|c|c|}
\hline 
Target property
& Predictors Set I &  Predictors Set II
                  &  Predictors Set III \\[5pt]
                  \hline
$\rm E_g~(HSE)$         & $\rm E_g~(PBE)$       & Elements Signatures      &      $\rm E_g~(PBE)$, Elements Signatures \\[5pt]%
\hline
$\rm VBM~(HSE)$         & $\rm VBM~(PBE)$       & Elements Signatures      &      $\rm VBM~(PBE)$, Elements Signatures \\ [5pt]%
\hline
$\rm CBM~(HSE)$         & $\rm CBM~(PBE)$       & Elements Signatures      &      $\rm CBM~(PBE)$, Elements Signatures\\[5pt]%
\hline
\end{tabular}
\label{table1}
\end{table*}

In addition to the type of machine learning methods we choose, a proper selection of feature space is also critical to achieve robust and accurate prediction. Previous studies usually include a large amount of predictors in the feature space and then conduct dimension reduction, which is likely to hide important physical insights of the model. Here, instead, we intend to compare the prediction power of PBE results as features and merely fundamental chemical and physical signatures of constituent elements in the materials. With this consideration, we have built three different sets of predictors: In Set-I, PBE results, band gap, VBM, and CBM, are the only feature used to predict related HSE values; In Set-II, we include only elemental signatures for each material, such as atomic mass, ionization energy, electron affinity, electronegativity, as well as electronegativity difference between cations and anions; Set-III is a combination of Set-I and Set-II. Features in Set-I depend on less time-consuming PBE calculations, while Set-II is more convenient to obtain with no requirement of any DFT calculations.

\subsection {Isoelectronic Materials Design}
\label{sec:mat_design}

\begin{figure}[t]
\includegraphics[width=\columnwidth]{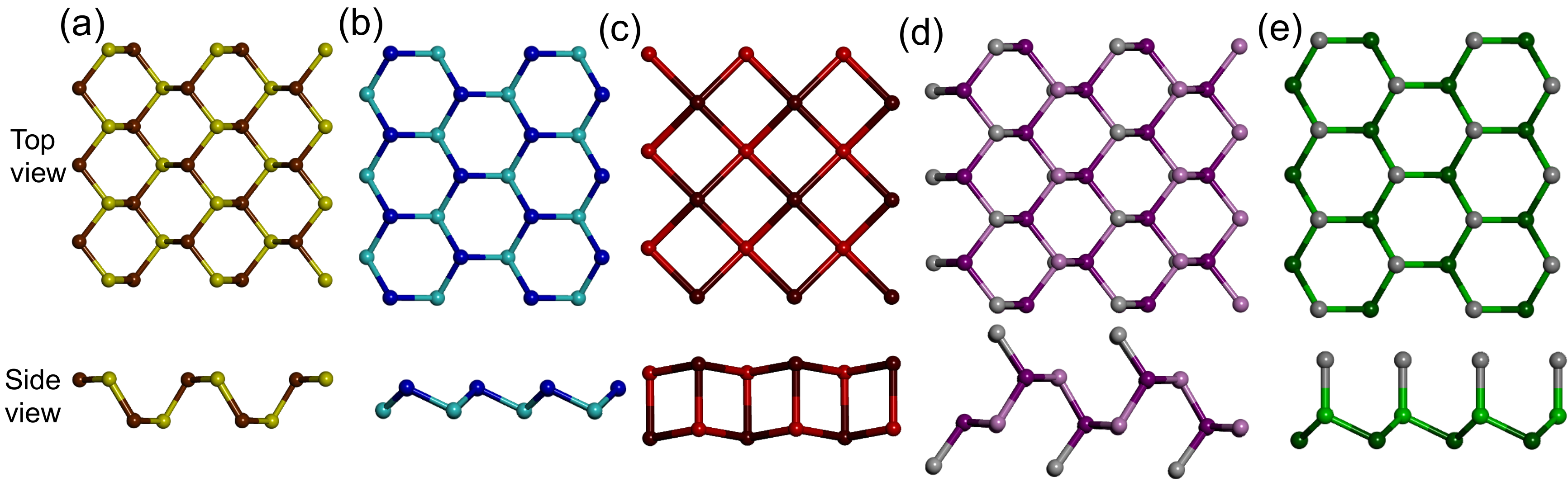}
\caption{(Color online) Equilibrium structures of elemental materials and binary compounds
in (a) Phase-I, (b) Phase-II, (c) Phase-III, and their ternary counterparts in (d) Phase-I and (e) Phase-II.
All structure are shown in both top and side views.
\label{fig:struc}}
\end{figure}

2D group-V elemental materials, such as phosphorene~\cite{DT229,Zhu2014BlueP} and antimonene~\cite{Zhang2015Sb}, can be stabilized in two distinct structural phases 
as shown in Fig.~\ref{fig:struc}(a) and (b). In both
structural phases, each atom forms three covalent bonds of $sp^3$ type with adjacent atoms, as well as lone-pair electrons, which fulfils the octet rule.
The pyramid formed by a center atom and its three nearest neighbors can be arranged in a variety of ways, thereby leading to a rich design space for structural polymorphs~\cite{Guan2014}.

Binary compounds can be derived from their elemental counterparts by cation mutation while the averaged valence electrons are conserved to be five~\cite{Zhu2015SiS}. Based on such a principle, group IV-VI and III-VII compounds can
be conveniently designed and they are isoelectronic to the well-studied group-V elemental materials. In addition to
two base structures mentioned previously, III-VII compounds can also be stabilized in a special structure with a primitive cell of approximately square shape, as shown in Fig.~\ref{fig:struc}(c). In fact, for Indium Iodide, an existing compound of the III-VII family, Phase-III is the most energetically favored structure among the polymorphs mentioned here~\cite{Wang2017}. Thus, we also include this structural phase as one of the base structure for 2D materials design in present study.

The isoelectronic design principle can be further generalized to construct ternary compounds. As shown in Fig.~\ref{fig:struc}(d) and (e), III-VI-VII and IV-V-VII compounds share similar structures as elemental and binary materials discussed above; indeed, they are isoelectronic. Taking phosphorene of Phase-II as the starting material, we change half of P atoms to a group IV element, such as Si. $sp^3$ bonding in the material is maintained and P atoms still have the close-shell electron configuration. However, as Si has one less valence electron, one unpaired electron exists for Si rather than a lone pair in P. Furthermore, a group-VII halogen element can form an additional bond with Si, thereby satisfying the octet rule for the ternary compounds. Therefore, IV-V-VII compounds are isoelectronic to group-V elemental materials. Similarly, III-VI-VII compounds can be shown as isoelectronic counterparts to IV-VI compounds. Importantly, in the element mutation process to construct the ternary compounds, half of the lone pairs in the original materials no longer exist, but instead form covalent bonds between the metal and halogen atoms. In fact, ternary compounds are not limited to these two groups of materials. Simply applying the cation-mutation principle to IV-VI and III-VII compounds, we can obtain III-V-VI$_2$ and II-IV-VII$_2$ ternary compounds. As they are expected to be rather similar to their parent binary compounds, these groups of ternary materials are not computed using DFT methods in present work, but instead their electronic properties can be predicted from our machine learning models that is to be discussed in Sec.~\ref{sec:disc}.

To build a database for this family of 2D materials, we have considered entire group-III, group-IV, group-V, group-VI, and group-VII elements (except the radiative Tl, Po, and At) for isoelectronic materials design. For elemental and binary materials, three structural phases, Phase-I, Phase-II, and Phase-III, are treated as the base structure to perform element mutation. Following the design principle above, we have constructed 15 elemental materials and 108 binary compounds. For ternary compounds, we only use Phase-I and Phase-II as the base to construct isoelectronic compounds, giving 328 distinct 2D materials. Then, we perform DFT calculations to obtain the optimized structures, the fundamental band gaps, and absolute positions of band edges at both PBE and HSE levels. In fact, not all the element combinations can maintain the structural phases we are interested in present work. Especially, materials
containing B, C, O, and N are in general not able to be stabilized in the desired structural form. The data points corresponding to these materials are eliminated from the database and not used for machine learning exploration.

\section {Electronic properties by DFT-PBE and HSE}
\label{sec:elec_prop}

\begin{figure}[t]
\includegraphics[width=\columnwidth]{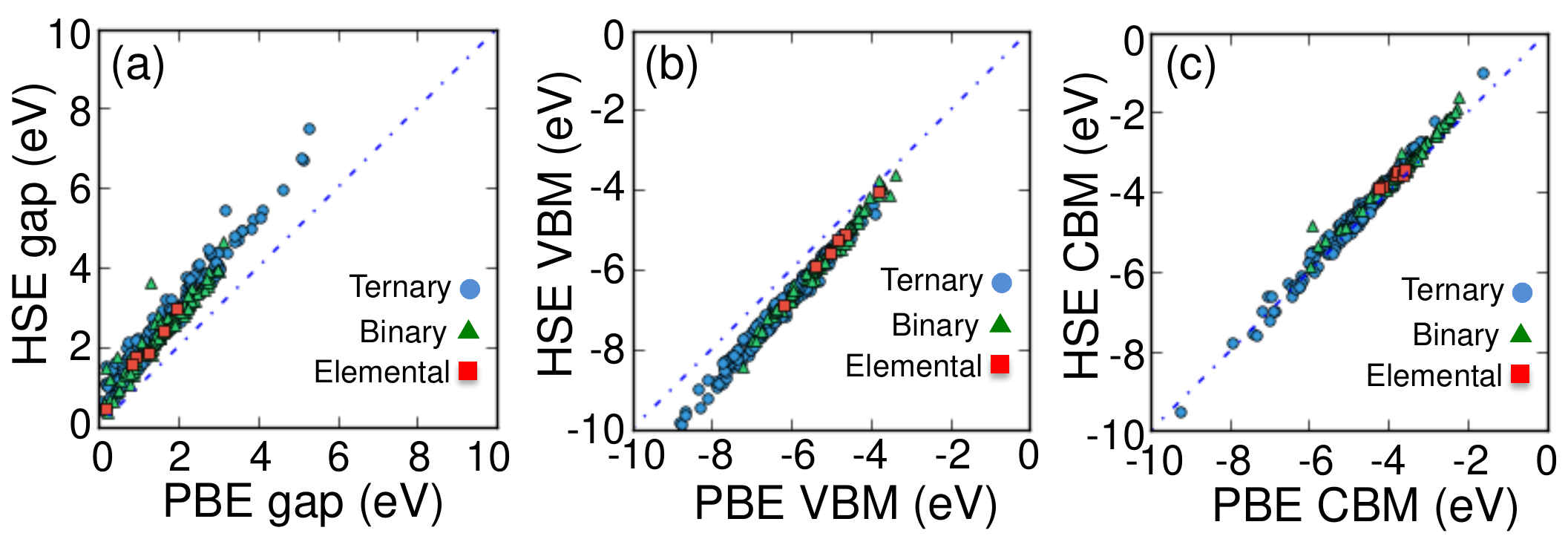}
\caption{(Color online) Relation between PBE and HSE results of (a) fundamental band gaps,
(b) VBMs, and (c) CBMs. The dashed lines are guide to the eye, indicating the case HSE values
equate PBE values. \label{fig:pbe_hse}}
\end{figure}

The calculated electronic properties: fundamental band gaps, VBMs, and CBMs, are shown in Fig.~\ref{fig:pbe_hse}.
It is known that for the $\it same$ semiconductor HSE band gap would scale linearly with the mixing parameter and DFT-PBE band gap value is in general the intercept. However, for $\it different$ materials, it is not clear how HSE band gaps are related to that predicted by DFT-PBE. Here in Fig.~\ref{fig:pbe_hse}, we have illustrated that
HSE band-gap values scale approximately linearly with that of DFT-PBE. The linear relation can be further improved when these isoelectronic materials are separated to different categories based on the number of constituent elements,
which is reflected by color-distinguished data points in Fig.~\ref{fig:pbe_hse}(a).

For absolute positions of band edges, the linear relationship between HSE and DFT-PBE is even more clear. The VBM position of a material, referenced to the vacuum level, corresponds to its electron ionization energy, which in general can be predicted by HSE to a good agreement with experiments. As shown in Fig.~\ref{fig:pbe_hse}(b), HSE
predicts lower VBMs than that of PBE and we also find a linear relationship between VBMs of HSE and PBE. Therefore, promisingly, PBE results may act as efficient descriptors for expensive HSE calculations, as well as experimental results, which is to be assessed in Sec.~\ref{sec:ml_model}. Similarly, HSE CBMs also scale linearly with that obtained by PBE, illustrated in Fig.~\ref{fig:pbe_hse}(c). However, for CBMs, HSE results are slightly higher than that of PBE, in sharp contrast to the case for VBMs.

\begin{figure}[b]
\includegraphics[width=\columnwidth]{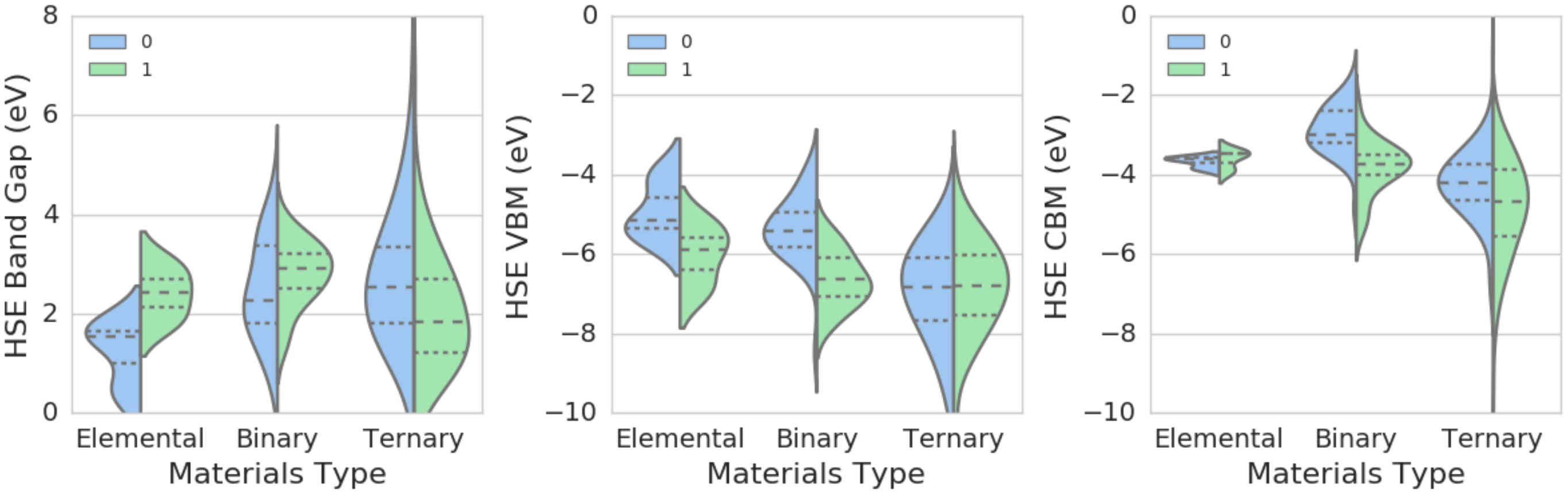}
\caption{(Color online) Distribution of (a) band gap values, (b) VBMs, and (c) CBMs based on HSE calculations for different materials types. The blue-colored areas represent materials of Phase-I, while the green-colored ones are for Phase-II. The long-dashed line indicates the mean of the distribution. \label{fig:gap_compare}}
\end{figure}

To gain deeper insights into electronic properties of this family of materials, we have shown the distribution of band gaps, VBMs, and CBMs with respect to both materials types and structural phases (Fig.~\ref{fig:gap_compare}). Clearly, for both elemental and binary materials, Phase-II structures have larger band gap values than that of Phase-I, which is closely related to the fact that VBMs of the former are in general lower
than that of the later, indicated in Fig.~\ref{fig:gap_compare}(b). The similar trend for these two groups of materials
can be explained by the fact that they are isoelectronic and the band-edge states are similar. On the contrary, for ternary compounds the averaged band gap value of Phase-I structures is larger than that of Phase-II structures by $\approx$1~eV; Especially, they have very similar distribution of VBMs. The different behaviors between ternary compounds from the others is due to the presence of halogen ligand that satisfied the octet rule without forming lone-pair electrons. In fact, ternary compounds are not ``perfectly'' isoelectronic to their parent compounds. Since the signature of long-pair electrons are still partially persevered in ternary compounds, CBMs share similar characters for all three types of materials (Fig.~\ref{fig:gap_compare}(c)). Furthermore, binary compounds are found to have larger averaged band gap than that of elemental materials, which can be attributed to the increased ionicity in the materials: a larger electronegativity difference between cation and anion usually leads to a larger band gap value~\cite{Goodman1958}. The factors that affect band gaps and alignments of materials are to be discussed in Sec.~\ref{sec:disc}.

\section {Machine learning predictive models}
\label{sec:ml_model}

As mentioned in Sec.~\ref{sec:dft_method}, computational results in present work are at two distinct levels of theory: DFT-PBE and the screened hybrid functional method (HSE). The former is computationally less demanding, but severely underestimates the fundamental band gap; HSE, on the other hand, can precisely predict band-gap values and alignments of standard semiconductors (without localized $d$ or $f$ orbitals as valence electrons), but is formidable for large-scale functional materials screening due to high computational cost. Therefore, it is desirable to build computational efficient methods that can also achieve high accuracy simultaneously. Given different predictor sets as described in Sec.~\ref{sec:ml_method}, we apply machine learning methods, including LR, RFR, and SVR, to predict computed electronic properties at HSE level, which can be further utilized to predict experimental observations.


\subsection{Set-I Predictors}
\label{sec:set_1}

\begin{figure}[t]
\includegraphics[width=\columnwidth]{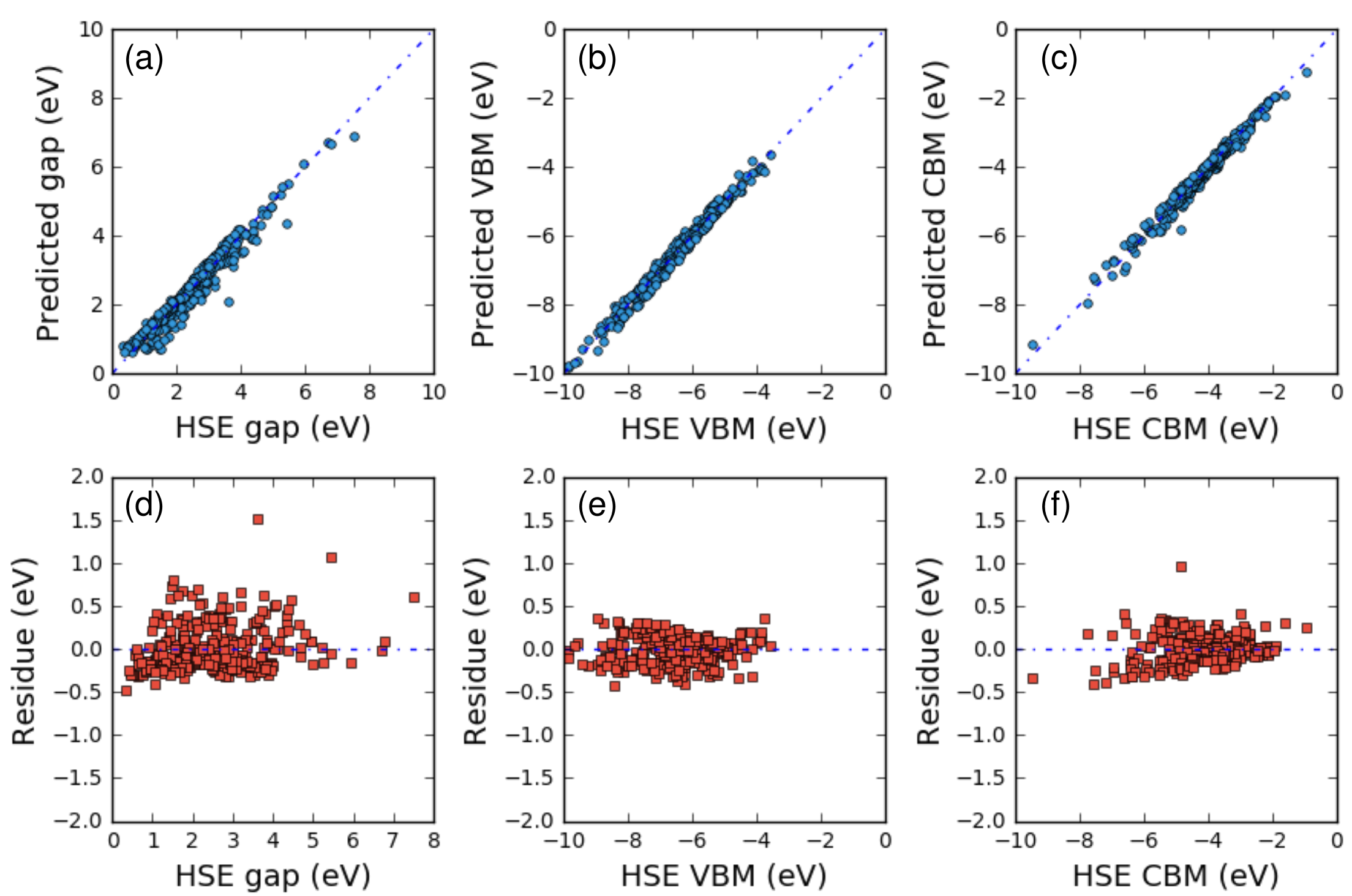}
\caption{(Color online) Comparison of predicted (a) band gaps, (b) VBMs, and (c) CBMs by LR with calculated HSE values.
Computed relevant PBE value (Predictors Set-I) is used as the single descriptor for the predictive LR model. Corresponding residues
are shown in (d), (e), and (f) to access the accuracy of the model. The dashed lines are guide to the eye, representing that predicted values are equal to computed HSE data.\label{fig:set_1}}
\end{figure}

As inferred from Sec.~\ref{sec:elec_prop}, HSE band gaps have approximately linear relation with that of PBE. Intuitively, PBE band gaps have been used as the only feature in Predictors Set-I. LR model is applied to model
the relation between results of HSE and PBE with 10-fold cross validation, while RFR and SVR are not applicable for
such simple feature space. The predicted HSE gaps of the validation
sets are shown with respect to calculated values, presented in Fig.~\ref{fig:set_1}(a). The relation is:
\begin{equation}
\rm {E}_g^{HSE}=1.21E_g^{PBE}+0.52~eV.
\label{eq:gap}
\end{equation}
The residues, difference between predicted and computed band-gap values, are presented in Fig.~\ref{fig:set_1}(d) and the majority fall into the [-0.5~eV,0.5~eV] energy range, indicating good prediction accuracy. There are only two data points where the difference is larger than 1.0~eV. Even though they might be outliers, their influence on our regression model is minimal. Furthermore, we also calculate the RMSE and MAPE to evaluate the predictive model
and the small prediction error, 0.25~eV for RMSE and 10.67\% for MAPE (see Table.~\ref{tab:error}), also reflects the high accuracy of the model.

Similarly, in order to predict VBM$^{\rm HSE}$ positions VBM$^{\rm PBE}$ values are used as the only feature in Set-I predictors space. The predicted VBM$^{\rm HSE}$ positions of validation sets are presented in Fig.~\ref{fig:set_1}(b), showing excellent agreement with targeted values. All the residues [Fig.~\ref{fig:set_1}(e)] are in the [-0.5~eV,0.5~eV] energy range. The better linearity of the VBM predictive model, comparing with that of band gap, also leads to smaller prediction errors as listed in Table.~\ref{tab:error}. The predicted relationship between VBM$^{\rm PBE}$ and VBM$^{\rm HSE}$ is:
\begin{equation}
\rm {VBM}^{HSE}=1.15VBM^{PBE}+0.23~eV.
\label{eq:vbm}
\end{equation}
CBM$^{\rm HSE}$ can also be predicted by CBM$^{\rm PBE}$ with LR method and model accuracy is illustrated in Fig.~\ref{fig:set_1}(c) and (f). For CBM, the relation between HSE and PBE results is:
\begin{equation}
\rm {CBM}^{HSE}=1.07CBM^{PBE}+0.51~eV.
\label{eq:cbm}
\end{equation}
Since these three linear models are cross-validated by randomly selected samples from our 2D materials data set, they should be universally valid for materials that are isoelectronic to current family members.

\begin{table*}[t]
\caption{Prediction errors for band gaps of the LR, RFR, and SVR models.}
\hspace{1cm}
\begin{tabular}{ll|cc|cc|cc}
\hline 
Regr. \quad
& Pred. \quad &  \quad Band Gap
                  & Band Gap \quad &   \quad VBM
                  &  VBM \quad &  \quad CBM
                  &  CBM\\
                Methods \quad 
& Sets \quad &  \quad RMSE (eV)
                  &  MAPE (\%) \quad &  \quad RMSE (eV)
                  &  MAPE (\%) \quad & \quad RMSE (eV)
                  &  MAPE (\%) \\[5pt]
                  \hline
         & Set-I       &   0.25   &   10.67 &   0.15   &    1.85 &  0.14    &   2.53    \\[5pt]%
$\rm LR$ & Set-II       &   0.87    &  35.07 &    0.88   &  10.30 &  0.80     &   16.03    \\[5pt]
& Set-III       &   0.15    &   5.55 &       0.09 &    1.04 &   0.09    &   1.56   \\[5pt]
\hline
& Set-I        & --      &      -- & --      &      -- & --      &      -- \\ [5pt]
$\rm RFR$         &   Set-II     &  0.70     &   26.37 &    0.67   &  7.23  &   0.57   & 10.22  \\[5pt]%
& Set-III       &  0.25     &   7.44 &  0.18     &    1.75  &  0.18     &    2.64   \\ [5pt]
\hline
& Set-I       & --      &      -- & --      &      -- & --      &      --\\[5pt]
$\rm SVR$         & Set-II       &   0.57    &  16.80 &   0.49    &  4.83 &   0.43    &  7.07 \\[5pt]
& Set-III       &     0.13  &   4.93 &     0.08  &   0.96 &     0.09  &   1.65\\[5pt]%
\hline
\end{tabular}
\label{tab:error}
\end{table*}




\subsection {Set-II Predictors}
\label{sec:set_2}

\begin{figure}[t]
\includegraphics[width=\columnwidth]{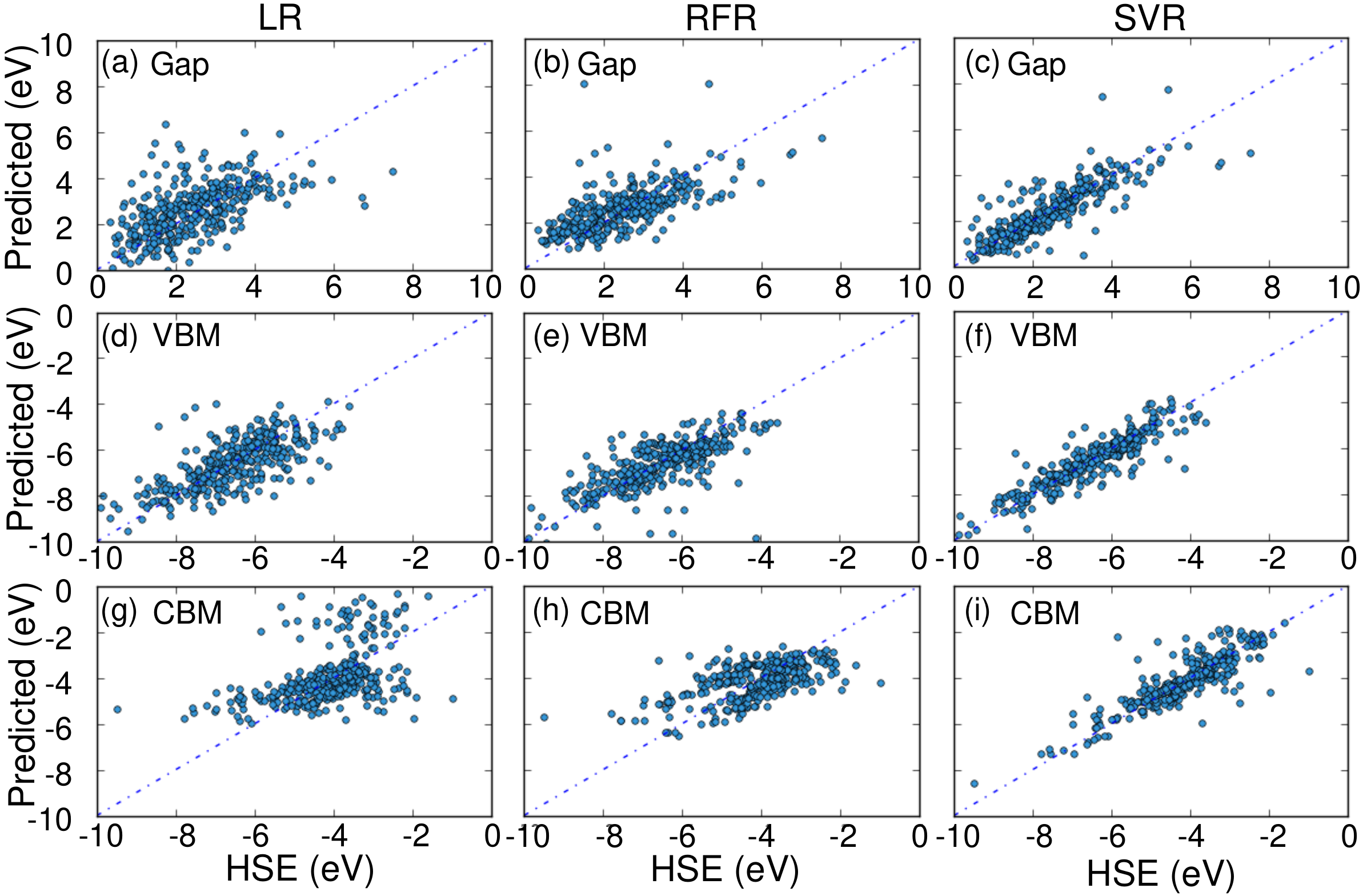}
\caption{(Color online)
Comparison of calculated HSE values with predicted fundamental band gaps and band edges by LR, RFR, and SVR models.
These models are evaluated with 10-fold cross-validation and only the
predicted results of the validation sets are shown. Each column of the subfigures
represents one predictive model and each row shows one selected electronic property.
Selected signatures of constituent elements are used as predictors (Predictors Set-II) for these
machine learning models.  The dashed lines are guide to
the eye, representing that predicted values are equal to computed HSE data.
\label{fig:set_2}}
\end{figure}

The ideal predictive model 
would rather have elemental information of constituent elements
as the feature space, instead of DFT results at any level of theory.
This would greatly improve the model efficiency and even make real-time interactive prediction possible. We have created Predictors Set-II to fulfill such a purpose. Details about this set of predictors are discussed in Sec.~\ref{sec:ml_method}.

For this set of predictors, we have applied LR, RFR, and SVR to predict the targeted electronic properties. The performance of these models are shown in Fig.~\ref{fig:set_2}, as inferred from the relationship between
predicted values and computed values.
Here the LR model shows much inferior predictive ability comparing with the case when PBE results are used as
predictors. This is also reflected by its high RMSE (0.87~eV) and high MAPE (35.07\%) as presented in Table.~\ref{tab:error}. Comparing with band-gap prediction, the accuracy of LR model is slightly improved for VBMs and CBMs: MAPE values are 10.30\% and 16.03\% respectively. To avoid overfitting, we have also compared simple LR model with regularized models, such as Ridge regression and LASSO, and found no improvement in the model performance.




The undesired performance of LR model indicates that nonlinear relationship between the Set-II predictors and
computed HSE results is essential. Complicated models, like RFR an SVR, are likely to capture the nonlinearity in the feature-target relation. Indeed, we find both RFR and SVR models have better performance than the former LR model.
SVR is found to give the lowest RMSEs: 0.57~eV for band gaps, 0.49~eV for VBMs, and 0.43~eV for CBMs, corresponding to MAPEs of 16.80\%, 4.83\%, and 7.07\%, which equate approximately 50\% error reduction from the LR model. Even though the performance is still inferior to LR with DFT-PBE results as features, it should be noted that the SVR model we developed here is of advantage to be used for fast materials screening due to its convenient feature space with no requirements for DFT calculations.

Although RFR is not the best predictive model, it can provide precious insights on important features that determines
underlying materials properties. Alongside training of a RFR model, we can also obtain the relative importance of predictors in the feature space. For band-gap prediction, the most significant feature is ``Average mass'': the heavier
the compounds, the smaller the band gap. It is noted that increased metallicity is inherited naturally from larger atomic mass for elements from same element group, which weakens both bonding strength and ionicity, result in the narrowing of band gap. Other important features are ``Electronegativity difference between cation and anion'', ``Cation electronegativity'', ``Phase type'', and so on. For VBMs and CBMs, the rankings of feature importance are different: ``Average mass'' is not as important as for band-gap prediction. VBMs depend strongly on ``Electronegativity difference between cation and anion'' while ``Anion electron affinity'' is the most significant factor determining CBMs.





\subsection {Set-III Predictors}
\label{sec:set_3}

\begin{figure}[t]
\includegraphics[width=\columnwidth]{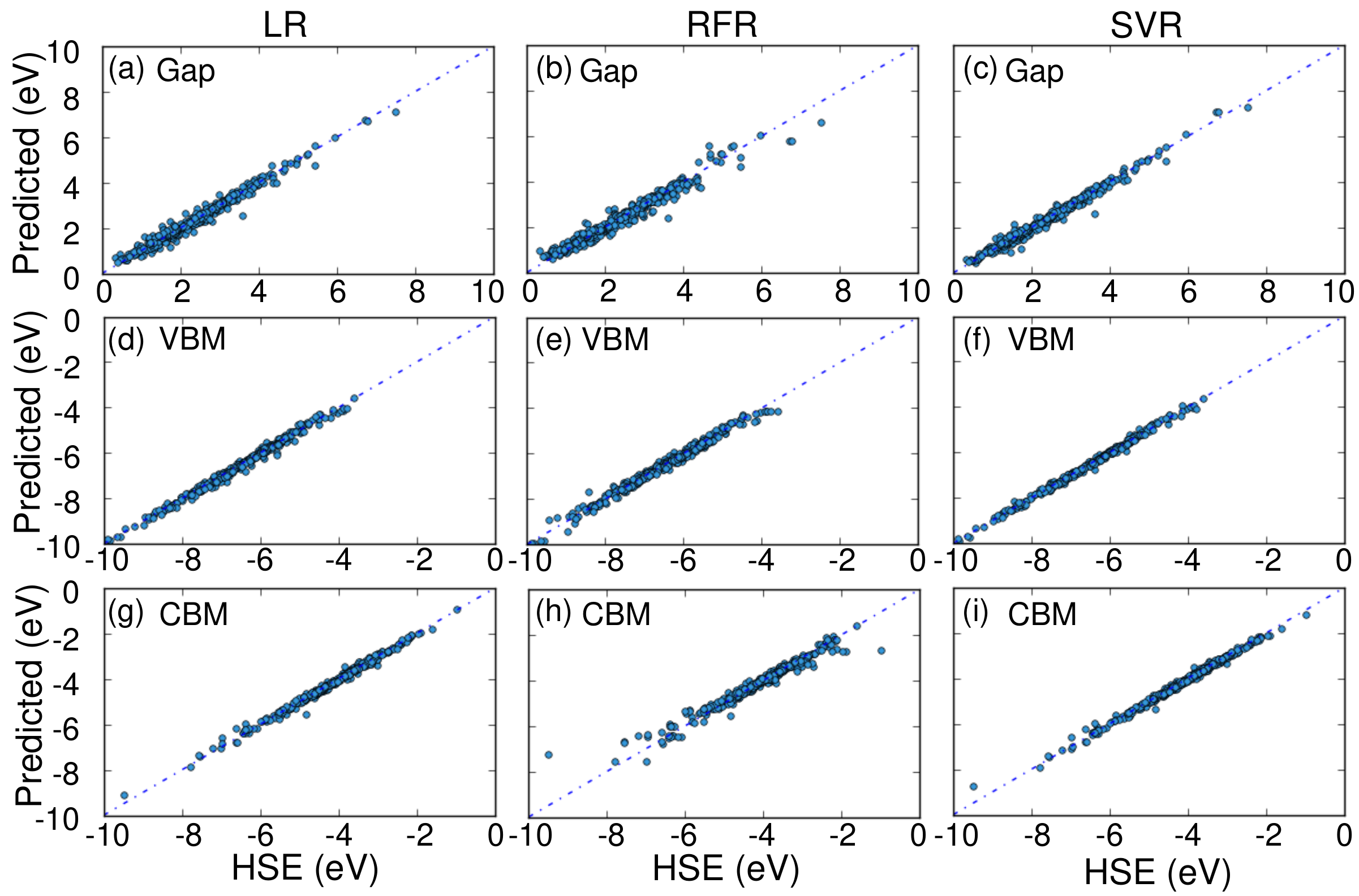}
\caption{(Color online) Comparison of calculated HSE values with predicted fundamental
band gaps and band alignments by LR, RFR, and SVR models, where both relevant PBE results
and selected signatures of constituent elements are used as predictors (Predictors Set-III).
These models are evaluated with 10-fold cross-validation and only the
predicted results of the validation sets are shown. Each column of the subfigures
represents one predictive model and each row shows one selected electronic property.
The dashed lines are guide to the eye, representing that predicted values are equal to
computed HSE data.\label{fig:set_3}}
\end{figure}

The predictive models can be further improved when predictors of Set-I and Set-II are combined as the
new feature space: Set-III Predictors. As DFT-PBE can also be viewed as a good predictive model, machine learning
methods based on Set-III Predictors can thus be viewed as a process of model stacking, which in general give better
prediction performance. The predicted results for validation sets are compared with computed values in Fig.~\ref{fig:set_3}. Indeed, we find the RMSE and MAPE of all three regression models are significantly reduced with
respect to the case where feature space is spanned by either Set-I or Set-II predictors. Among the machine learning models used here, SVR outperforms the other two for all three targeted materials properties, with RMSEs of 0.13~eV for band gap, 0.08~eV for VBM, and 0.09~eV for CBM. In fact, the prediction errors are within the accuracy of HSE calculations, thereby justifying the validity and accuracy of our models for properties prediction.



\subsection {Discussions}
\label{sec:disc}

{\bf Model selection}. By carefully comparing the performance of different machine learning models, we have elucidate the general principle for model selection.
Both RMSE and MAPE are computed to evaluate the accuracy of LR, RFR, and SVR models. Among these three models in present study, SVR is found to have the best accuracy, when either Set-II or Set-III predictors are used as the feature space. Especially, when only elemental information is used to span the feature space, SVR show significant advantage in predicting targeted materials properties over other two methods. Therefore, SVR is suggested to use when no prior DFT-PBE results are available. On the other hand, if DFT-PBE values are available, LR is a good model to start with. In this method, the relation between HSE values and DFT-PBE features can be expressed in a simple analytical model, thus target values can be readily predicted.

{\bf Performance for different target properties.} Same machine learning model is found to have different performance when target properties vary. Even though band gap is closely related to VBM and CBM, the later two targets almost always have smaller RMSE and MAPE than the former. The difference in accuracy is likely caused by the fact that
for band-gap prediction a good predictor reflecting both VBM and CBM states is a requisite, which is unlikely to be included in our simple feature space. On the other hand, for VBM- or CBM-prediction the requirement is less stringent and more likely to be covered by our selection of predictors. To further improve model performance, we expect to have a more complicated feature space, including different operations between predictors in current feature space. This is beyond the scope of current study.

{\bf Applications}. As mentioned in Sec.~\ref{sec:mat_design}, materials used in present work 
are just a small fraction of this large family of materials following proposed isoelectronic materials design principle. Our trained models, especially SVR with Set-II Predictors, can be applied to predict fundamental band gaps and alignments of other family members with minimal computation cost. The predicted results are informative and valuable even when
the designed materials 
are not the most stable structural phase. It has been shown that alloying unstable materials with stable ones in the desired structural phase is likely to stabilize the former compounds. For example, CaSe can be stabilized in Phase-I when alloying with SnSe~\cite{Matthews2017}. The electronic properties of such alloys can also be predicted by our models where the weighted average of constituent elements are taken as predictors. Therefore, the trained machine learning models in present study provide a computational efficient method to accurately obtain band gaps and alignments of a large amount of 2D materials, which enables fast screening of 2D functional materials for electronic, optoelectronic, and photocatalysis applications.

\section {Conclusions}
\label{sec:summary}

In conclusion,
we have explored fundamental band gaps
and alignments of a group of two-dimensional
semiconductors isoelectronic to phosphorene using machine learning techniques
in combination with density functional theory. This family of
2D materials shares similar crystalline structures, but possesses
unprecedented rich band-gap values
ranging approximately from 0 to 8~eV.
Based on machine learning methods, we
trained predictive models
that can predict band-gap values and band-edge positions with surprisingly high accuracy.
Among models discussed in present work, SVR is found to have the best performance with RMSEs less than 0.15~eV for predicted band gaps, VBMs, and CBMs
when both PBE results and elemental information are used as predictors.
We also demonstrate the predictive models can be utilized for electronic properties prediction
for more complicated systems, like quaternary compounds and alloys, shedding light on rational
materials design for (opto)electronic and photocatalysis applications.

\section {Acknowledgements}
This work is dedicated to Michelle Mucheng Zhu. This study was supported by the National Basic Research
Program (No.2017YFA0206301) of China and the Major Program of Aerospace
Advanced Manufacturing Technology Research Foundation NSFC and CASC, China
(No. U1537204). Computational resources have been provided by High Performance Computing Center
of the Institute of Metal Research, Chinese Academy of Sciences.



\end{document}